# Correlated Electrons and Magnetism in Double Perovskites

GAYANATH W. FERNANDO*

*Department of Physics, University of Connecticut, Storrs, CT06269, USA*

SAIKAT BANERJEE

*Institute of Physics, University of Greifswald, Felix-Hausdorff-Strasse 6, 17489 Greifswald, Germany*

R. MATTHIAS GEILHUFE

*Department of Physics, Chalmers University of Technology, 412 96 Göteborg, Sweden*

This paper is an overview of some recent work done on the double perovskites. We discuss the physics of selected double perovskite compounds emphasizing the relevant interactions and resulting observable phenomena such as the magnetic order using different theoretical approaches. Spin-Orbit interaction, which is comparable to other relevant interaction strengths, plays a central role in determining the physics of such *4d-5d* perovskites.

*Keywords*: Double Perovskites, Magnetism, Correlated Electrons, Density Functional Theory

## 1. Introduction

Double perovskites have numerous unusual but useful properties. They are suitable for applications in optics, electronics and photonics. In addition, they are highly active in catalysis for oxygen evolution and there are potential applications in thermoelectrics and photovoltaics such as in solar cells. In the green energy-related technologies, perovskite-based solar cells have been extensively examined as an alternative for Si-based ones. Another advantage is that they are non-toxic compared to certain Pb-based single perovskites.

In some of these perovskites, strong electron correlations combined with Jahn-Teller effects and spin-orbit coupling (SOC) can bring out fascinating and perhaps hidden orders. SOC is central to several phenomena such as spin or anomalous Hall effect, Rashba coupling and skyrmion physics.

In this overview, we will explore possible microscopic models relevant to heavy 4d/5d Mott insulators, summarize some of the related work reported in the past and compare our results obtained using state-of-the-art density functional theory (DFT) calculations as discussed below. Typically, the correlated aspects in these materials are described within the multi-orbital Hubbard-Kanamori Hamiltonian (Refs. 1 and 2).

---

* gayanath.fernando@uconn.edu

In contrast, weak hopping between the orbitals are usually captured by the Slater-Koster integrals (Ref. 3). The study of transition metal (TM) oxides/halides in this context goes back to the early days of many-body physics; however, it is still quite active because of its enormous potential for yet unexplored novel phases and functionalities of quantum matter. Double perovskite materials hosting electronic configurations $d^1 - d^6$ are usually Mott insulators. In heavy transition metal-based oxides, SOC strength is comparable to other energy scales (such as exchange-correlation) and hence spin and orbital angular momenta are not individually conserved. Instead, we must focus on the total angular momentum *J* which determines various ground state and other properties, such as the multi-polar orders present in the material. However, some of such orders could be hidden as suggested in some experimental constructions of their phase diagrams. In addition, strong correlations are expected to play a significant role here.

The following questions will be examined in this review on double perovskites:
- How does the spin-orbit interaction affect the physics of these compounds?
- How does the Jahn-Teller type symmetry breaking affect physics?
- What role does exchange interaction play here?
- What about strong correlations?

As mentioned earlier, there have been many previous studies on transition metal-based double perovskites (Refs. 4-15). Early on, Kugel and Khomskii (Ref. 4) discussed magnetic insulators containing orbitally degenerate transition metal ions. More recently, one of us (RMG) has looked into composite quadrupole order in ferroic and multiferroic materials in a theoretical study (Ref. 15). In the compounds discussed here there are possible multipolar orders present at higher temperatures. The DFT calculations presented below are exclusively for the T=0 ground state indicating possible dipolar order only. In these, we assign a certain *(constrained)* dipolar order in the starting magnetic configuration and monitor its evolution through the self-consistent electronic and structural energy minimizations. Recently, more sophisticated approaches have been proposed using a constrained initialization of the on-site density matrix derived from a *multipolar ordered* effective *ab initio* Hamiltonian as reported in Ref. 16.

### 1.1. An example of a single perovskite TiF$_3$ (with weak spin-orbit coupling)

From ours as well as other studies (such as reflectivity measurements), we concluded that the single perovskite TiF$_3$ is an insulator having moderate correlations. However, the spin-orbit parameter associated with Ti ion is at least ten times smaller than those for the 5d atoms Re and Os considered here. Hence it is not expected to play a significant role in determining magnetic order. Non-collinear magnetic order appears to be present in the ground state which was evident from noncollinear VASP calculations as well as a double-exchange Hubbard model-based theoretical work. A detailed discussion of this work can be found in Ref. 17.

**1.2. Double Perovskites (with strong spin-orbit coupling)**

The formula unit for the double perovskites of interest is $A_2BB'O_6$ where $A$ is Barium or a similar atom, $B$ is a non-magnetic atom occupying opposite corners of a cube and $B'$ is a magnetic atom occupying the other corners of the cube. For example, in this paper, we will examine two double perovskites (i) $Ba_2NaOsO_6$ or (ii) $Ba_2MgReO_6$. The following figure illustrates the cubic geometry and $d_{xy}$ type orbitals with the TM positions identified as B and B'.

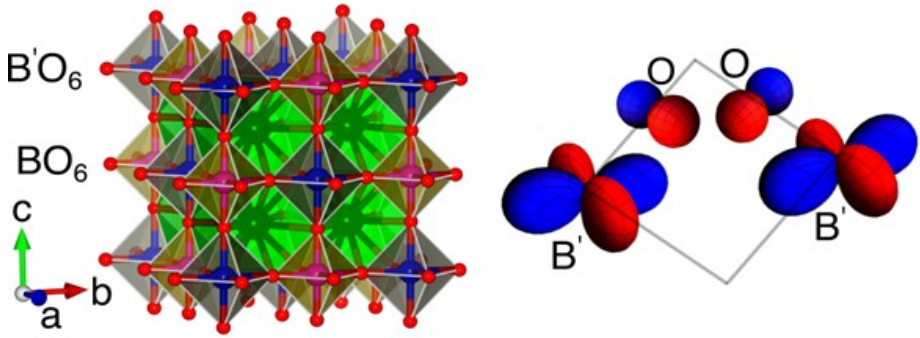

Fig. 1: (Left) A double perovskite structure $A_2BB'O_6$ in cubic geometry is illustrated with Oxygen atoms depicted by small (red) circles. (Right) TM $d_{xy}$ type and Oxygen $p$-type orbitals showing possible exchange paths for the TM B' atoms through Oxygen.

We will focus mostly on the cubic $d^1$ and $d^2$ double perovskites here. Under cubic symmetry, due to octahedral crystal field splitting, $e_g$ levels lie well above the $t_{2g}$ levels. In addition, the $d$-orbital $t_{2g}$ sector is 3-fold degenerate and isomorphic to the $p$-orbital sector with an effective angular momentum of **L**= -1. This can be shown from a direct operator equivalence between the $d$-electron $t_{2g}$ matrix elements and the $p$-electron matrix elements (using their native basis sets, see Ref. 13). This **L**$_{eff}$ = -1 added to spin (for $d^1$ electronic configurations) gives rise to a total *local* moment of **L**$_{eff}$ + **2S** leading to zero net moment for an isolated ion. In reality, there is a multiplicative factor κ in **L**, called the covalency factor, which actually leaves the above sum close to zero. Due to this, the net moment remains low but non-zero (as we will see later from our calculations). When strong spin-orbit interaction is present, **L** and **S** are no longer conserved, and the ground state is found to be in a **J** = **L**+**S** state.

The situation is markedly different for $4d^2/5d^2$ transition metal ions. In these cases, a finite magnetic moment arises from the combined effect of the effective orbital angular momentum (**L**$_{eff}$) and twice the spin (**2S**), mediated by Hund's coupling between the two electrons. Analogous effects can also be envisioned for $d^4$ and $d^5$ electronic configurations, corresponding to two and one hole(s), respectively, in the $t_{2g}$ shell. Furthermore, strong spin-orbit coupling in all these configurations leads to a splitting of the degenerate atomic $t_{2g}$ levels into multiplets characterized by the total angular momentum $\mathbf{J} = \mathbf{L} + \mathbf{S}$, as schematically illustrated in Fig. 2. In our consideration, we specifically ignore materials with $d^3$ electronic electronic configurations. This half-filled $t_{2g}$ case is interesting as the total angular momentum of the atomic state is quenched due to strong Hund's coupling and to the zeroth order does not host any effective spin-orbit coupling. However, there could be virtual processes which might trigger various level splitting dependent on the hierarchy of the Hund's coupling, spin orbit coupling and the crystal field.

We now turn our attention to *ab initio* results relevant for the $d^1$ configuration in specific double perovskite compounds. Before proceeding, it is helpful to highlight a few key features identified in related systems studied by us and others in recent years. A complementary case is that of $d^5$ Mott insulators, which can be regarded as the one-hole perovskite materials like the iridates ($A_2IrO_3$, A = Na, Li) and α-$RuCl_3$, both of which have been proposed as candidate Kitaev quantum spin liquid systems. In these cases, the low-energy atomic manifold is a J = 1/2 Kramers doublet, and the crystal structure typically features *edge-shared octahedra*, which plays a crucial role in mediating anisotropic exchange interactions along [111] directions in the crystal (cleavage plane, see Refs. 25-30). A typical magnetic exchange Hamiltonian in this case reads as

$$\mathcal{H} = J \sum_{\langle ij \rangle} \mathbf{S}_i \cdot \mathbf{S}_j + K \sum_{\langle ij \rangle, \gamma} S_i^\gamma S_j^\gamma + \Gamma \sum_{\langle ij \rangle, \alpha\beta} \left( S_i^\alpha S_j^\beta + S_i^\beta S_j^\alpha \right) + \sum_i (\mathbf{S}_i \cdot \mathbf{n})^2$$

where $(\alpha,\beta,\gamma)$ represents a permutation of $(x, y, z)$, the local coordinates of the transition metal octahedron, and $\langle ij \rangle$ corresponds to nearest-neighbor sites in a honeycomb lattice along [111] plane. The Heisenberg exchange interaction $J$ denotes generic Heisenberg exchange coupling, K represents the bond dependent Kitaev coupling, $\Gamma$ denotes the symmetric off-diagonal exchange interactions, and $A$ corresponds to the single-ion anisotropy along direction $\mathbf{n}$. In practical aspect, almost all of the candidate materials show magnetically ordered state at low temperatures. One could tune the relative strength of the couplings by external drive such as circularly polarized light, and fine tune the parameter space to achieve the ideal limit with (K $\gg$ $J, \Gamma, A$) relevant for Kitaev spin liquid phase (see Ref. 27). Note that in the ideal limit, the spins are extremely frustrated and does not host any magnetically ordered state.

The situation becomes even more intricate for $d^2$ and $d^4$ electronic configurations. In both cases, strong Hund's coupling favors a total spin $\mathbf{S} = 1$, corresponding to two electrons (in the $d^2$ case) or two holes (in the $d^4$ case) occupying the $t_{2g}$ shell. The addition of strong spin-orbit coupling further yields a total angular momentum $\mathbf{J} = \mathbf{L} + \mathbf{S} = 2$. Depending on whether one adopts the electron or hole perspective, the resulting low-energy atomic multiplet can manifest as either a $J_z = 2$ quintuplet or a $J_z = 0$ singlet -- each leading to qualitatively distinct physics. In the $d^2$ case, the J = 2 quintuplet can further split under a cubic crystal field into an effective $T_{2g}$ (triplet) and $E_g$ (non-Kramers doublet), analogous to crystal field splitting of an isolated *d*-electron atom (see Refs. 18, 30). If the $E_g$ doublet lies lowest, it can support higher-order magnetic multipoles, including quadrupole and octupole moments, specifically devoid of dipole moments. One of us recently showed that such multipolar moments can induce ferroelectric polarization [Ref. 18]. Most importantly, such multipolar order can in principle couple to the relevant Jahn-Teller modes and should be a viable way to test them in real materials using experimental probes, or *ab initio* methods.

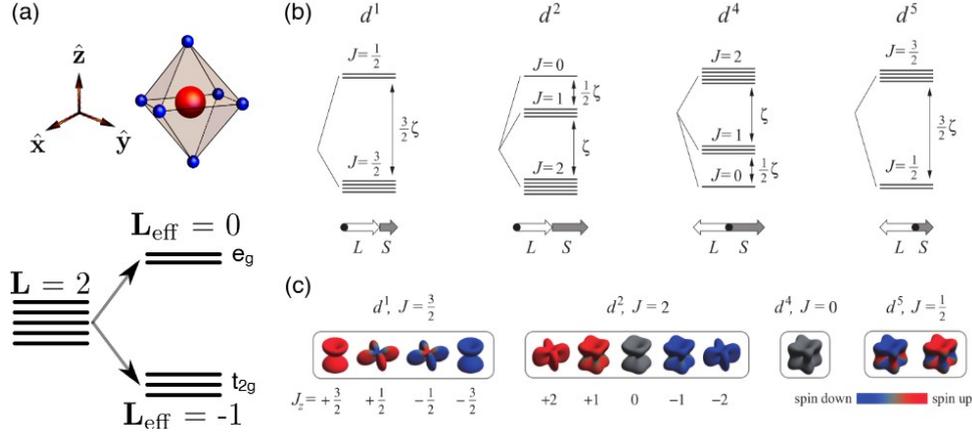

Fig 2: *J*-orbital splittings and corresponding $J_z$ orbital shapes for $d^1$, $d^2$, $d^4$ and $d^5$ configurations in the presence of strong spin-orbit coupling (from Ref. 6). For electron counts less than half-filled in the $t_{2g}$ shell, spin-orbit coupling aligns the effective orbital angular momentum ($\mathbf{L}_{\text{eff}}$) and spin ($\mathbf{S}$), resulting in a larger total angular momentum. Specifically, the $d^1$ configuration gives rise to a $J = 3/2$ quartet, while the $d^2$ configuration yields a $J = 2$ quintuplet ground state. In contrast, for more-than-half-filled $t_{2g}$ shells, $\mathbf{L}_{\text{eff}}$ and $\mathbf{S}$ are anti-aligned. This leads to a $J = 0$ singlet ground state in the $d^4$ configuration, whereas the $d^5$ configuration hosts a pseudospin $J = 1/2$ Kramers doublet.

In contrast, $d^4$ Mott insulators generically host a $\mathbf{J} = 0$ singlet ground state with a gapped $\mathbf{J} = 1$ triplet excitation, as proposed by G. Khaliullin. This leads to intriguing possibilities such as excitonic magnetism, where condensation of the excited triplet can give rise to novel magnetic orders (see Ref. 31). It will be interesting to test these propositions using state-of-the-art *ab initio* calculations, which we plan to explore in future.

## 2. Density Functional Calculations on Double Perovskites

In the following, we present results from non-collinear first principles calculations on the two Double Perovskites (a) $Ba_2MgReO_6$ and (b) $Ba_2NaOsO_6$. Here the spin-orbit interaction strength (of Re and Os) is relatively high compared to Ti in $TiF_3$ case. These two double perovskites show somewhat contrasting behavior with respect to their magnetic properties.

The magnetic atom B' (Fig. 1) in the formula unit occupies a face-centered cubic lattice, and the magnetic ordering may be neither FM nor AFM. There are several effects at play here, including strong spin-orbit coupling, $t_{2g}$ - $e_g$ splitting, and *j=3/2* states within a $d^1$ configuration.

We have allowed the possibility of non-collinear magnetism here in order to probe non-traditional magnetic order. These self-consistent calculations were carried out within the framework of the density functional theory (Refs. 19-20), as implemented in the Vienna Ab Initio Simulation Package (VASP) (Ref. 21). The exchange-correlation functional was approximated by the generalized gradient approximation (GGA) (Ref. 22), with the energy cutoff chosen according to potential input files. For integration in *k*-space, a 6x6x6 (triclinic cell) Γ-centered mesh according to Monkhorst and Pack (Ref. 23) was used during each self-consistent cycle. Structural optimization was performed until the Hellman-Feynman (Ref. 24) forces acting on the atoms were negligible. *In most DFT calculations reported so far on similar systems, structural optimization is not carried out as we have done here. There are non-negligible movements of Oxygen atoms because of optimization and the breaking of cubic symmetry is clearly observed.* A unit cell consisting of twice (and four times) the size of the formula unit was also used here so that it is possible to allow self-consistent variations of the angle between the moments of the Re atoms in the unit cell.

2.1.1. *Distortions away from the cubic structure*

In VASP, when Hellman-Feynman forces are used to displace the atoms during the self-consistent electronic structure iterations, atoms will be displaced from their starting cubic positions. These displacements could be used as a measure of the Jahn-Teller effect that is present in these materials. Here we allow variations of atomic positions as well and computed an index tied to these displacements in all the structures. One notable feature observed at self-consistency is that the transition metal atom positions maintain their cubic positions while the Oxygen atoms get displaced. Below is a table that shows the change in fractional coordinates found in these compounds, separated into contributions from all the atoms and contributions from only Oxygen atoms in a 20-atom unit cell. From this table it is evident that at least 80% of the distortions arise from Oxygen atoms, indicating their Jahn-Teller-type effects in symmetry breaking.

Table 1. A listing of the distortions away from the starting cubic structure observed in the fully relaxed VASP calculations in fractional coordinates for a 20-atom cell as well as the distortions only for Oxygen atoms. These are the expected symmetry breaking Jahn-Teller distortions.

| Compound | $\Delta Q$ (all atoms) | $\Delta Q$ (Oxygen) |
| --- | --- | --- |
| $Ba_2ReMgO_6$ | 0.269 | 0.223 |
| $Ba_2OsNaO_6$ | 0.394 | 0.316 |
| $Ba_2ReCaO_6$ | 0.366 | 0.299 |

**2.2.**

2.2.1. *Density of States*

One of the most straightforward ways to determine if the calculated structure is a metal or an insulator is to compute its Density of States (DOS) for the chosen unit cell. Here we illustrate two-unit cell sizes, 20 and 40 atoms for $Ba_4Na_2Os_2O_{12}$ It is noteworthy that a VASP calculation with U= 0.0 eV, J= 0.0 eV produces metallic behavior, indicating that correlations play a central role in the insulating behavior of the compound. Similarly, for the other double perovskites examined here, metallic behavior is seen in the absence of correlations. A moderate U value of about 3 eVs produces a band gap of about 0.3 eV (Fig. 3), as observed in experiments.

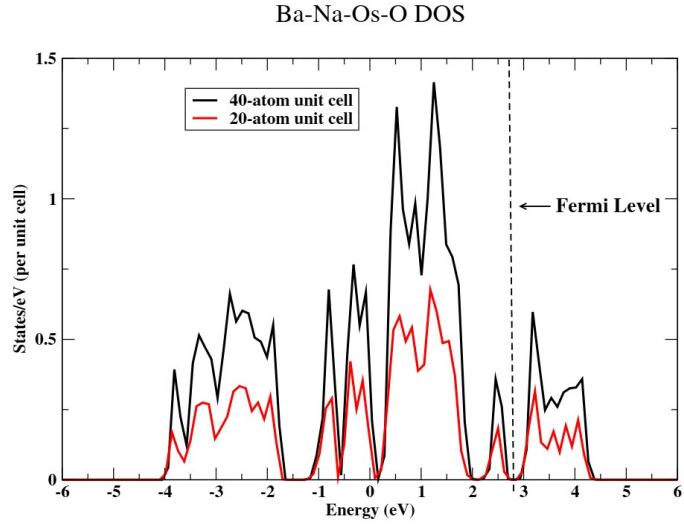

Fig: 3: Density of States for the double perovskite $Ba_2OsNaO_6$ using 20 and 40 atom unit cells as discussed in the text. There is a gap of about 0.3 eV in both plots at the Fermi level indicating insulating behavior for the chosen parameter values U=3 eV and J=0.5 eV. For U=0 and J=0 eV, there is no gap, indicating the effect of correlations.

### 2.2.2. *Total Energy and Magnetic Moments*

Total energies in VASP calculations are outcomes from a self-consistent procedure involving electronic and ionic degrees of freedom and they provide a reliable estimate of a possible global equilibrium configuration. In addition, these are tied to a well-controlled variational scheme where relaxations are allowed using Hellman-Feynman forces. In the following, we have evaluated total energies while varying the angle between the moments of the two magnetic atoms in a 20-atom unit cell of the compounds $Ba_4Os_2Na_2O_{16}$ and $Ba_2ReMgO_8$. The former compound indicates possible ferromagnetic canting as evident from the figure while the latter shows possible antiferromagnetic order (as seen in Figs. 4 and 5).

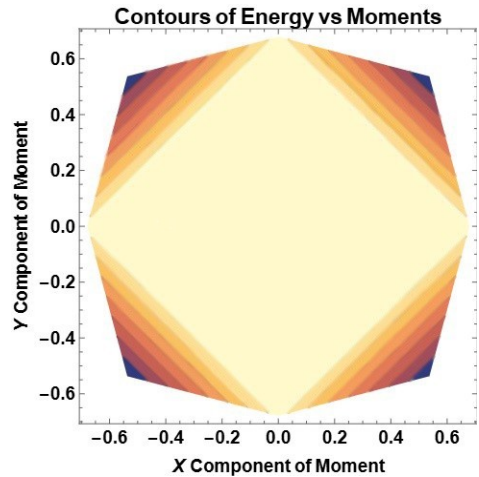

Fig. 4: A false color plot of ferromagnetic moments in $Ba_4Os_2Na_2O_{16}$ where the X and Y axes labels are in units of Bohr Magnetons. Darker colors indicate the regions where energy is lower for the given XY cubic directions (as in $d_{xy}$-type orbital directions). The darker colors also identify the regions where actual calculations produced the indicated ferromagnetic moment. A very similar behavior is seen for XZ and YZ directions. The center region shows nonmagnetic (and metallic) behavior. The magnetic regions are clearly insulating and the self consistently calculated Os moments agree with experiment. These results also point to possible ferromagnetic canting.

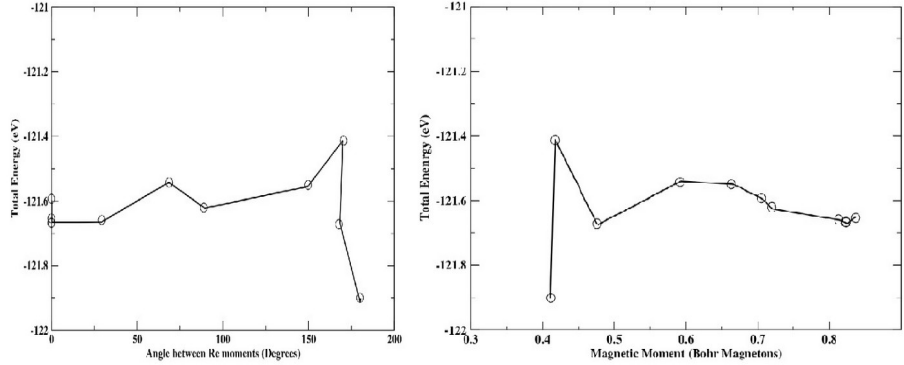

Fig 5: Total energy vs angle between Re moments and total energy vs magnetic moment in the compound. $Ba_2ReMgO_8$. These plots show that the magnetic moments are low in this compound as expected from a *J=3/2* state (discussed previously).

### 2.2.3. **Summary**:

In this article, we have briefly reviewed the current status of the theoretical aspects of the physics of the double perovskite compounds emphasizing the relevant interactions such as octahedral crystal field energy, onsite Coulomb repulsions, Hund's coupling and spin-orbit interaction. We briefly touched upon various resulting observable phenomena such as the novel magnetic, or multipolar ordering, Jahn-Teller effect using different theoretical approaches. Spin-orbit interaction which is comparable to other interaction strengths plays a central role in determining the physics of such *4d-5d* perovskites. In conjunction with Hund's coupling and the crystal field energy, the spin-orbit coupling leads to various low-energy multiplet structures varying over $d^1$ - $d^6$ electronic configurations as illustrated in Fig. 2(b). We also performed zero temperature ab initio calculations for possible magnetic ordering in $d^1$ systems, and discussed the potential role of Jahn-Teller distortions in driving the novel magnetic ordering. We stayed away from any potential aspects of multipolar ordering since we focused on zero temperature analysis, however, plan to perform more detailed analysis regarding the interaction between magnetic, multipolar ordering and their coupling to the specific Jahn-Teller modes in a future publication.

**Acknowledgements:** We gratefully acknowledge the assistance of Dr. Deyu Lu at Brookhaven National Laboratory, Center for Functional Nanomaterials (CFN), for carrying out some VASP6 calculations reported here. In addition, we acknowledge the

**References**

(1) J. Kanmori, Prog. Theo. Phys. 17,177(1957).
(2) J. Kanmori, Prog. Theo. Phys. 17, 197(1957)
(3) J. C. Slater and G. F. Koster, Phys. Rev 94, 1498 (1957).
(4) K. Kugel and D. Khomskii, Sov. Phys. Usp. 25, 231 (1982).
(5) J. Romhanyi, L. Balents, and G. Jackeli, 118, 217202 (2017).
(6) T. Takayama, J. Chaloupka, A. Smerald, G. Khaliullin, and H. Takagi, J. Phys. Soc. Japan 90, 062001 (2021).
(7) S. Gangopadhyay, and W. E. Pickett, Phys. Rev. B 91, 045133 (2015).
(8) S. V. Streltsov, and D. I. Khomskii, Phys. Rev. X 10, 031043 (2020).
(9) G. Chen, R. P ereira, and L. Balents, Phys. Rev. B 82, 17440 (2010).
(10) G. Chen, and L. Balents, Phys. Rev. B 84, 094420 (2011).
(11) L. Lu, M. Song, W. Liu, A. P. Reyes, P. Kuhns, H. O. Lee, I. R. Fisher, and V. F. Mitrovic, Nat. Comms. 8:14407, DOI:10.1038.
(12) W. Witczak-Krempa, G. Chen, Y. B. Kim, and L. Balents, Annu. Rev. Condens. Matter Phys. 5:57-82 (2014).
(13) G. L. Stamokostas, and G. A. Fiete, Phys. Rev. B 97, 085150 (2018).
(14) L. V. Pourovski, D. F. Mosca, and C. Franchini, Phys. Rev. Lett. 127, 237201 (2021).
(15) R. M. Geilhufe, J. Phys.: Condens. Matter, 37 05LT01.
(16) D. F. Mosca, L. V. Pourovskii and C. Franchini, Phys. Rev. B **106**, 035127 (2022)
(17) G. W. Fernando, D. Sheets, J. Hancock, A. Ernst, and R. M. Geilhufe, Phys. Status Solidi Rapid Res. Lett., 2300330 (2023).
(18) S. Banerjee, S. Humeniuk, A. R. Bishop, A. Saxena, and A. V. Balatsky, Phys. Rev. B **111**, L201107 (2025)
(19) P. Hohenberg and W. Kohn, Phys. Rev. **136**, B864 (1964).
(20) W. Kohn and L. J. Sham, Phys. Rev. **140**, A1133 (1965).
(21) G. Kresse and D. Joubert, Phys. Rev. B **59**, 1758 (1999).
(22) J. P. Perdew, K. Burke, and M. Ernzerhof, Phys. Rev. Lett. **77**, 3865 (1996).
(23) J. H. Monkhorst and J. D. Pack, Phys. Rev. B **13**, 5188 (1976).
(24) R. P. Feynman, Phys. Rev. **56**, 340 (1936)
(25) J. Chaloupka, G. Jackeli and G. Khaliullin, Phys. Rev. Lett. **105**, 027204 (2010)
(26) J. G. Rau, E. K.-H. Lee and H.-Y. Kee, Phys. Rev. Lett. **112**, 077204 (2014)
(27) U. Kumar, S. Banerjee, and S.-Z. Lin; Commun. Phys. **5**, 157 (2022)
(28) S. Banerjee, U. Kumar, and S.-Z. Lin; Phys. Rev. B **105**, L180414 (2022)
(29) S. Banerjee, and S.-Z. Lin; SciPost Phys. **14**, 127 (2023)
(30) G. Khaliullin, D. Churchill, P. P. Stavropoulos, and H.-Y. Kee, Phys. Rev. Research **3**, 033163 (2021)
(31) G. Khaliullin, Phys. Rev. Lett. **111**, 197201 (2013)